\documentclass[journal]{IEEEtran}

\ifCLASSINFOpdf
\else
\fi

\usepackage{dsfont}
\usepackage{amsmath,hhline}
\usepackage{color}
\usepackage{multirow}
\usepackage[T1]{fontenc}
\usepackage{url,cite}
\usepackage{lastpage,algorithm,algorithmic,graphicx}
\usepackage{epsfig,amssymb,caption}
\usepackage{bbding}

\IEEEoverridecommandlockouts 

\begin{document}
\vspace{-4cm}
\title{Context-Aware Mobility Management in HetNets: A Reinforcement Learning Approach\thanks{This research was supported in part by the SHARING project under the Finland grant 128010 and by the U.S. National Science Foundation
under the grants CNS-1406968 and AST-1443999.}}

\author{\IEEEauthorblockN{Meryem Simsek$^*$, Mehdi Bennis$^\dag$,  and \.{I}smail G\"uven\c{c}$^\ddag$\\}
\IEEEauthorblockA{$^*$Vodafone Chair Mobile Communications Systems
Technische Universit\"at Dresden, Germany\\
$^\dag$Centre for Wireless Communications, University of Oulu, Finland\\
$^\ddag$Department of Electrical \& Computer Engineering, Florida International University, USA\\
Email: meryem.simsek@tu-dresden.de,  bennis@ee.oulu.fi, and iguvenc@fiu.edu}
}
\maketitle

\begin{abstract}

The use of small cell deployments in heterogeneous network (HetNet) environments is expected to be a key feature of 4G networks and beyond, and essential for providing higher user throughput and cell-edge coverage. 
However, due to different coverage sizes of macro and pico base stations (BSs), such a paradigm shift  introduces additional requirements and challenges in dense networks. Among these challenges is the handover performance of user equipment (UEs), which will be impacted especially when high velocity UEs traverse picocells. In this paper, we propose a coordination-based and context-aware mobility management (MM) procedure for small cell networks  using tools from reinforcement learning. Here, macro and pico BSs jointly learn their long-term traffic loads and optimal cell range expansion,  and schedule their UEs based on their  velocities and historical rates (exchanged among tiers). The proposed approach is shown to not only outperform the classical MM in terms of UE throughput, but also to enable better fairness. In average, a gain of up to 80\% is achieved for UE throughput, while the handover failure probability is reduced up to a factor of three  by the proposed learning based MM approaches.
\end{abstract}

\section{Introduction}
The deployment of Long Term Evolution (LTE) heterogeneous networks (HetNets) is a promising approach to meet the ever-increasing wireless broadband capacity challenge~\cite{cisco,Damnjanovic}. However, deploying  HetNets entails a number of challenges in terms of capacity, coverage, mobility management (MM), and mobility load balancing (MLB) across multiple network tiers~\cite{3GPP36839}. Mobility management is essential to ensure a continuous connectivity to mobile user equipment (UEs)  while maintaining quality of service (QoS). 

The mobility framework for LTE was originally developed and  analyzed by the $3^{\text{rd}}$ generation partnership project (3GPP) for  macro-only networks, and was therefore not explicitly optimized for HetNets. In LTE Rel.~11, \emph{mobility enhancements in HetNets} have been investigated through a dedicated study item~\cite{3GPP36839}. 3GPP has defined key performance indicators (KPIs) for mobility measurements, i.e., the handover failure (HOF) due to a degraded signal-to-interference-plus-noise-ratio (SINR),
the radio link failure (RLF), as well as the probability of unnecessary handovers, typically referred to as
ping-pong (PP) events. 

Poor MM approaches may increase the HOFs, RLFs, and PPs, and result in unbalanced load among cells. This entails a low resource utilization efficiency and hence deterioration of the user experience. In order to solve this problem, while minimizing PPs, mobility parameters in each cell need to be carefully and dynamically optimized according to cell traffic loads. It is essential to optimize handover parameters such as time to trigger (TTT), range expansion bias (REB), and hysteresis margin in order to answer the question: ``\emph{when} to handover \emph{which} UE to \emph{which} cell?''

Mobility management techniques  for HetNets have been recently investigated in the literature, e.g., in~\cite{Barbera,David,Feki,Wang}. In~\cite{Barbera}, the authors evaluate the effect of different combinations of MM parameter settings for HetNets. The main result is that mobility performance strongly depends on the cell size and UE speed. The simulations in~\cite{Barbera} consider that all UEs have the same velocity in each simulation setup.
In ~\cite{David}, the authors evaluate the mobility performance of HetNets considering almost blank subframes in the presence of cell range expansion and propose a mobility based intercell interference coordination (ICIC) scheme. Hereby, picocells configure coordinated resources by muting certain subframes so that macrocells can schedule their high velocity UEs in these resources without co-channel interference from picocells. However, the proposed approach only considers three broad classes of UE velocities: low, medium, and high. Moreover, no adaptation of the REB has been taken into account. 
In~\cite{Feki}, a handover-aware ICIC approach based on reinforcement learning is proposed. Hereby, the authors model the ICIC approach as a sub-band selection problem for mobility robustness optimization in a small cell only network.  
In~\cite{Wang}, the cell selection problem in HetNets is formulated as a network wide proportional fairness optimization problem by jointly considering the long-term channel condition and load balance in a HetNet. While the proposed method enhances the cell-edge UE performance, no results related to mobility parameters are presented.

To the best of our knowledge there is no previous work related to learning based mobility management in HetNets by jointly considering load balancing and UE scheduling. In this paper, we propose a joint MM and context-aware UE scheduling approach by using tools from reinforcement learning. Hereby, each base station (BS) individually optimizes its own strategy (REB, UE scheduling) based on limited coordination among tiers. Both macro- and picocells learn how to optimize their traffic load in the long-term and  the UE association process in the short-term by performing history and velocity based scheduling. We propose  multi armed bandit (MAB) and satisfaction based MM learning approaches aiming at improving the overall system performance and reducing the HOF  and PP probabilities. 

\begin{figure}[t]
	\centering
	\includegraphics[width=0.43\textwidth]{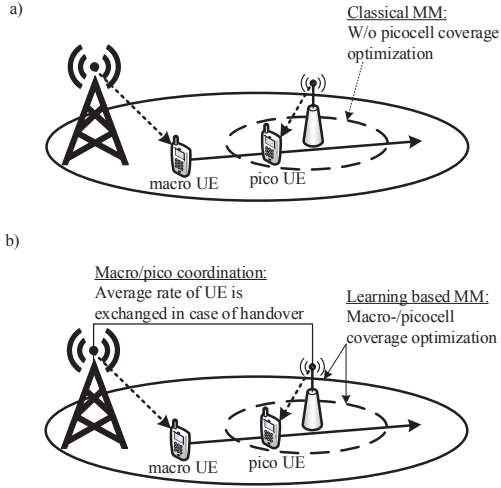}
	\caption{a) Classical MM framework w/o picocell coverage optimization, b) Proposed learning based MM framework considering velocity and history (average rate) based scheduling.}
	\label{fig:MMframework}
\end{figure}
To illustrate the differences between the classical MM and our proposed approach, we depict in Fig. \ref{fig:MMframework} a) and Fig. \ref{fig:MMframework} b) the basic idea of the classical MM and proposed MM approaches, respectively. In the classical MM approach, there is no information exchange among tiers in case of UE handover and traffic offloading might be achieved by picocell range expansion. In the proposed MM approaches, instead, each cell individually optimizes its own MM strategy based on limited coordination among tiers. The major difference between MAB and satisfaction based learning is that MAB aims at maximizing the overall capacity while satisfaction based learning aims at satisfying the network in terms of capacity. In both cases, macro and pico BSs learn on the long-term how to optimize their REB, which results in load-balancing. On the short-term, based on these optimized REB values, each cell carries out user scheduling by considering each UE's velocity and average rate, through coordinated effort among the tiers. Our contributions are as follows:
\begin{itemize}
\item In the proposed MM approaches, we focus on both short-term and long-term solutions. In the long-term, a traffic load balancing procedure in a HetNet scenario is proposed, while in the short-term the UE association process is solved. 
\item To implement the long-term load balancing method, we propose two learning based MM approaches by using reinforcement learning techniques: a MAB based and a satisfaction based MM approach.
\item The short-term UE association process is based on a proposed context-aware scheduler considering a UE's throughput history and velocity to enable fair scheduling and enhanced cell association.
\end{itemize}

The rest of the paper is organized as follows. Section~II describes the system model, the problem formulation for MM, and the context-aware scheduler. In Section~III, we introduce the learning based MM approaches. Section~IV  presents system level simulation results, and finally, Section~V concludes the paper.

\section{System Model}
We focus on the downlink transmission of a 2-layer HetNet, where layer 1 is modeled as macrocells and layer 2 as picocells. The HetNet consists of a set of BSs $\mathcal{K} = \{1,\ldots,K\}$ with a set $\mathcal M = \{1,\ldots,M\}$ of macrocells underlaid  by  a set $\mathcal P = \{1,...,P\}$ of picocells, where $\mathcal{K} = \mathcal{M} \cup \mathcal{P}$. Macro BSs are dropped following a hexagonal layout including three sectors. Within each macro sector $m$, $p\in \mathcal{P}$ picocells are randomly positioned, and a set $\mathcal U = \{1,...,U\}$ of UEs which are randomly dropped within a circle around each picocell $p$ (hotspot). The UEs associated to macrocells are referred as macro UEs $\mathcal{U}(m) = \{1(m),\ldots,U(m)\} \in \mathcal{U}$ and the UEs served by picocells are referred as pico UEs $\mathcal{U}(p) = \{1(p),\ldots,U(p)\} \in \mathcal{U}$, where $\mathcal{U}(p) \neq \mathcal{U}(m)$. Each UE $i(k)$ with $k\in\{m,p\}$ has a randomly selected velocity $v_{i(k)}\in\mathcal{V}$ km/h and a random direction of movement within an angle of $[0;2\pi]$. A co-channel deployment is considered, in which picocells and macrocells operate in a system with a bandwidth $B$ consisting of $r = \{1, \ldots, R\}$ resource blocks (RBs).
At every time instant $t_n = n T_{\rm s}$ with $n=[1,\ldots,N]$ and {$T_{\rm s}  = 1\text{ ms}$}, each BS $k$ decides how to expand its coverage area by learning its REB $\beta_k = \{\beta_m,\beta_p\}$ with $\beta_m=\{0;3;6\}$~dB and $\beta_p=\{0;3;6;9;12;15;18\}$~dB\footnote{We consider lower REB values for macro BSs to avoid overloaded macrocells due to their large transmission power.}. Both macro and pico BSs select their REB to decide which UE $i(k)$ to schedule on which RB based on the UE's context parameters. These context parameters are defined as the UE's velocity $v_{i(k)}$, its instantaneous rate $\phi_{i(k)}(t_n)$ when associated to BS $k$ and its average rate $\overline{\phi}_{i(k)}(t_n)$ defined as $\overline{\phi}_{i(k)}(t_n) = \frac{1}{T} \sum_{n=1}^N \phi_{i(k)}(t_n),$
whereby $T = N T_{\rm s} $ is a time window. The instantaneous rate $\phi_{i(k)}(t_n)$ is given by:
\begin{align}\label{eq:InstRate}
\phi_{i(k)}(t_n) &= B_{i(k)} \cdot \log\left(1+\gamma_{i(k)}(t_n)\right),
\end{align}
with $\gamma_{i(k)}(t_n)$ being the SINR of UE $i(k)$ at time $t_n$, which is defined as:
\begin{equation}
\gamma_{i(k)}(t_n) = \dfrac{p_k\cdot g_{i(k),k}(t_n)}{\sum\limits_{\substack{j\in\mathcal{K}\\j\neq k}}p_j \cdot g_{i(k),j}(t_n)+\sigma^2},
\end{equation}
with  $p_k$ being the transmit power of BS $k$, and $g_{i(k),k} (t_n)$ being the channel gain from cell $k$ to UE $i(k)$ associated to BS $k$. The bandwidth $B_{i(k)}$ in equation \eqref{eq:InstRate} is the bandwidth which is allocated to UE $i(k)$ by BS $k$ at time $t_n$. 

\subsection{Handover Procedure}\label{sec:HOprocedure}
According to the 3GPP standard, the handover mechanism is based on RSRP measurements, the filtering of measured RSRP samples, Handover Hysteresis Margin, and TTT mechanisms~\cite{3GPP36839}. 
A handover is executed if the target cell's (biased) RSRP (plus hysteresis margin) is larger than the source cell's (biased) RSRP. In summary, the handover condition for a UE $i(k)$  to BS $k$ is defined as:
\begin{align}\label{eq:HOcondition}
& P_{l}(i(l))  + \beta_l  < P_{k}(i(k))   + \beta_k   + m_{\text{hist}},
\end{align}
with $\{l,k\}\in \mathcal{K}$,  $ m_{\text{hist}}$ is the UE- or cell-specific hysteresis margin, $\beta_k (\beta_l)$ is the REB of BS $k (l)$, and $P_{k}(i(k))\left( \text{or } P_{l}(i(l))\right)$ [dBm] is the $i(k)$-th ( or $i(l)$-th) UE's RSRP from BS $k (l)$ after TTT.

\vspace{-0.3cm}
\subsection{Problem Formulation}\label{sec:ProblemFormulation}
Our optimization approach aims at maximizing the total rate of the network. Hereby, we consider long-term and short-term processes. The long-term load balancing optimization approach is solved by the proposed learning based MM approaches presented in Section \ref{sec:MAB} and Section \ref{sec:Satisfaction}, which result in REB $\beta_k$ value optimization and in load balancing $\phi_{k,\text{tot}}(t_n)$. Based on the estimated instantaneous  load, the context-aware scheduler selects, in the short-term, for each RB a UE by considering its history and velocity as described in Section \ref{sec:Scheduler}. This results in each UE's instantaneous rate $\phi_{i(k)}(t_n)$ and the RB allocation vector ${\boldsymbol{\alpha}_{i(k)}(t_n) = \left[\alpha_{i(k),1},...,\alpha_{i(k),R} \right]}$ containing binary variables $\alpha_{i(k),r}$, and indicating whether UE $i(k)$ of BS $k$ is allocated at RB $r$ or not. At each time instant $t_n$, each BS $k$ performs the following optimization:

\vspace{-0.3cm}
\begin{align}\label{eq:ProblemFormulation}
&\max_{\substack{\boldsymbol{\alpha}_{i(k)}(t_n)\\\beta_k}} \sum_{n=1}^N \sum_{{i}(k)\in \mathcal{U}_k} \sum_{r=1}^R \alpha_{i(k),r}(t_n) \cdot \phi_{i(k),r}(t_n) \\\nonumber
&\text{subject to:} \\
&\alpha_{i(k),r}(t_n) \in \{0,1\} \\
&\sum_{{i}(k)\in \mathcal{U}_k} \alpha_{i(k),r} = 1 \hspace{0.3cm} \forall r, \forall k,\\
&p_{k} \leq p_{k}^{\text{max}}\\
&\phi_{i(k)}(t_n) \geq \phi_{k,\text{min}},
\end{align}

\noindent where $\phi_{i(k),r}(t_n)$ is the instantaneous rate of UE $i(k)$ at RB $r$. The condition in (7) implies that the total transmitted power over all RBs does not exceed the maximum transmission power $p_{k}^{\text{max}}$ of BS $k$.  
\vspace{-0.3cm}
\subsection{Context-Aware Scheduler}\label{sec:Scheduler}
The proposed MM approach does not only optimize the load according to Section~\ref{sec:ProblemFormulation}, but considers also context-aware and fairness based UE scheduling. At each RB $r$, a UE $i(k)$ is selected to be served by BS $k$ on RB $r$ according to the following scheduling criterion:

\begin{align}\label{eq:scheduling}
&{i(k)_r}^{*} = \begin{matrix}
 \text{sort} \\
 \min{(v_{i(k)})}
\end{matrix}\left(\arg\max_{i(k)\in \mathcal{U}_k} \dfrac{\phi_{i(k),r}(t_n)}{\overline{\phi}_{i}(t_n)}\right),
\end{align}

\noindent where  $\text{sort}_{\min(v_{i(k)})}$ sorts the candidate UEs according to their velocity starting with the slowest UE, i.e. if more than one UE can be selected for RB $r$, the UE with minimum velocity is selected. The rationale behind introducing the sorting/ranking function for candidate UEs according to their velocity is that high-velocity UEs will not be favored over slow moving UEs. 

A scheduler according to \eqref{eq:scheduling} will allocate many (or even all) resources to a newly handed over UE since its average rate $\overline{\phi}_{i}(t_n)$ in the target cell is zero, i.e. in the classical Proportional Fair scheduler, $\overline{\phi}_{i}(t_n) = \overline{\phi}_{i(k)}(t_n) = 0$ when a UE is handed over to cell $k$, whereas we redefine it according to \eqref{eq:AvRate}. To avoid this and enable a fair resource allocation among all UEs in a cell, we propose a history based scheduling approach. We define the average rate $\overline{\phi}_{i}(t_n)$ according to \eqref{eq:AvRate} incorporating the following idea: Via the X2-interface macro- and picocells coordinate, so that once a macro UE $i(m)$ is handed over to picocell $p$ its rate history  at time instant $t_n$ is provided to picocell $p$ in terms of average rate $\overline{\phi}_{i(m)}(t_n)$, such that the UE's (which is named as $i(p)$ after the handover) average rate at picocell $p$ becomes:
\vspace{-0.1cm}
\begin{align}\label{eq:AvRate}
\overline{\phi}_{i(p)}(t_n+T_{\rm s}) = \frac{T\cdot\overline{\phi}_{i(m)}(t_n) + \phi_{i(p)}(t_n+T_{\rm s})}{T+T_{\rm s} }.
\end{align}
In \eqref{eq:AvRate}, a moving average rate is considered from macrocell to picocell, whereas in the classical MM approaches a UE's history is not considered and is equal to zero.  In other words, the proposed MM approach considers the historical rate when UE $i(m)$ was associated to the macrocell $m$ in the past. 
\vspace{-0.3cm}
\section{Learning Based Mobility Management Algorithm}\label{sec:Learning}
To solve the optimization approach defined in Section \ref{sec:ProblemFormulation}, we rely on the self organizing capabilities of HetNets and propose an autonomous solution for load balancing by using tools from reinforcement learning~\cite{Harmon}. Hereby, each cell develops its own MM strategy to perform optimal load balancing based on the proposed learning based approaches presented in Section~\ref{sec:MAB} and Section~\ref{sec:Satisfaction}.  To realize this, we consider the game $\mathcal{G}=\{\mathcal{K},\{\mathcal{A}_k\}_{k\in\mathcal{K}},\{u_k\}_{k\in\mathcal{K}}\}$. Hereby, the set $\mathcal{K}=\{\mathcal{M}\cup\mathcal{P}\}$ represents the set of players (i.e., BSs), and for all $k\in\mathcal{K}$, the set $\mathcal{A}_k = \{\beta_k\}$ represents the set of actions player $k$ can adopt. For all $k\in \mathcal{K}$, the function $u_k(t_n)$ is the utility function of player $k$. The players learn at each time instant $t_n$ to optimize the load in long-term and to perform context aware scheduling in short-term based on the algorithms presented in Section \ref{sec:MAB} and \ref{sec:Satisfaction} by the following steps:
\begin{enumerate}
\item Action $a_k\in\mathcal{A}_k$ is selected based on the obtained utility $u_k(t_n) = \phi_{k,\text{tot}}(t_n)$ with $\phi_{k,\text{tot}}(t_n)$ being the total rate of player $k$ at time $t_n$ as defined in equation \eqref{eq:load}.
\item The action selection strategy is updated based on the selected learning algorithm presented in Section \ref{sec:MAB} and Section \ref{sec:Satisfaction}.
\item UE of BS $k$ is allocated at RB $r$ based on its velocity, its instantaneous rate, and its average rate according to \eqref{eq:scheduling}.
\end{enumerate}

\subsection{Multi-Armed Bandit Based Learning Approach}\label{sec:MAB}
The objective of the MAB approach is to maximize the overall system performance. MAB is a machine learning technique based on an analogy with the traditional slot machine (one armed bandit)~\cite{Auer}. When pulled at time $t_n$, each machine/player provides a reward. The objective is to maximize the collected reward through iterative pulls, i.e. learning iterations. The player selects its actions based on a decision function reflecting the well-known exploration-exploitation trade-off in learning algorithms.


The set of players, actions and the utility function for our MAB based MM approach is  defined as follows:

\begin{itemize}
\item \textbf{Players:} Macro BSs $\mathcal{M}=\{1,\ldots,M\}$ and pico BSs $\mathcal{P}=\{1,\ldots,P\}$.
\item \textbf{Actions:}  $\mathcal{A}_k=\{\beta_{k}\}$ with
$\beta_{m} = [0, 3, 6]$ dB and $\beta_{p} = [0, 3, 6, 9, 12, 15, 18]$ dB being the CRE bias. We consider higher bias values for picocells due to their low transmit power. The considered bias values rely partially on the assumptions in~\cite{3GPPR1-113806} and at the same time extensive simulation results.
\end{itemize}

\begin{itemize}
\item \textbf{Strategy:}
\begin{enumerate}
\item Every BS learns its optimum CRE bias value on a long-term basis considering its load: 
\begin{align}\label{eq:load}
&{\phi}_{k,\text{tot}}(t_n) = \sum_{{i}(k)\in \mathcal{U}_k}\sum_{r=1}^R \alpha_{i(k),r}(t_n) \cdot \phi_{i(k),r}(t_n).
\end{align}
This is inter-related with the handover triggering by defining the cell border of each cell,
\item A UE is handed over to  BS $k$ if it fulfills the condition \eqref{eq:HOcondition}.
\item RB based scheduling is performed based on equation \eqref{eq:scheduling}.
\end{enumerate}
\item \textbf{Utility Function:}
The utility function in MAB learning is a decision function composed by an exploitation term represented by player $k$'s total rate and exploration part considering the number of times an action has been selected so far. Player $k$ selects its action ${a_{j(k)}(t_n)\in \mathcal{A}_k}$ at time $t_n$ through maximizing a decision function $d_{k,a_{j(k)}}(t_n)$, which is defined as:
\vspace{-0.3cm}
\begin{equation}\label{eq:decision}
d_{k,a_{j(k)}}(t_n) = u_{k, a_{j(k)}}(t_n) + \sqrt{\dfrac{2\log\left(\sum_{i=1}^{|\mathcal{A}_k|} n_{k,a_{i(k)}}(t_n)\right)}{n_{k,a_{j(k)}}(t_n)}},
\end{equation}
whereby $u_{k, a_{j(k)}}(t_n)$ is the mean reward of player $k$ at time $t_n$ for action $a_{j(k)}$, $n_{k,a_{j(k)}}(t_n)$ is the number of times action $a_{j(k)}$ has been selected by player $k$ until time $t_n$, and $|\cdot|$ represents the cardinality. 
\end{itemize}

During the first $t_n=|\mathcal{A}_k|\cdot T_{\rm s}$ player $k$ selects each action once in a random order to initialize the learning process by receiving a reward for each action. For the following iterations ${t_n>|\mathcal{A}_k|\cdot T_{\rm s} }$  action selection is performed according to Algorithm \ref{alg:MAB}. In each learning iteration the action $a^*_{j(k)}$ that  maximizes the decision function in \eqref{eq:decision} is selected. Then the parameters are updated, whereby the following notation is used: $s_{k,a_{j(k)}}(t_n)$ is the cumulated reward of player $k$ after playing action $a_{j(k)}$ and $\mathds{1}_{i=j}$ is equal to 1 if $i = j$ and zero otherwise.
\begin{algorithm}[t]
	\caption{MAB based mobility management algorithm.}
\begin{algorithmic}[1]
 \FOR{$t_n$}
\FOR{$i = 1:|\mathcal{A}_k|$}
\STATE Select action $a_{j(k)}^*$ according:
\STATE $a_{j(k)}^*=\arg\max_{a_{j(k)}\in |\mathcal{A}^k|}\left({d_{k,a_{j(k)}}(t_n)}\right)$
\STATE Update parameters according to:
\STATE Update the cumulated reward when player $k$ selects action $a_{j(k)}$
\STATE $s_{k,a_{j(k)}}(t_n+T_{\rm s} ) = s_{k,a_{j(k)}}(t_n)+ \mathds{1}_{i=j}\cdot \phi_{k,\text{tot}}(t_n)$
\STATE $n_{k,a_{j(k)}}(t_n+T_{\rm s} ) = n_{k,a_{j(k)}}(t_n) + \mathds{1}_{i=j}$
\STATE $u_{k,a_{j(k)}}(t_n+T_{\rm s} ) = \frac{s_{k,a_{j(k)}}(t_n+T_{\rm s} )}{n_{k,a_{j(k)}}(t_n+T_{\rm s} )}$
\ENDFOR
\STATE $t_n=t_n+T_{\rm s} $
\ENDFOR
   \end{algorithmic}
	\label{alg:MAB}
\end{algorithm}
\vspace{-0.5cm}
\subsection{Satisfaction Based Learning Approach}\label{sec:Satisfaction}
Satisfaction based learning approaches guarantee to satisfy the players in a system~\cite{Ross}. 
Here, we consider the player to be satisfied if its cell reaches a certain minimum level of total rate and if at least 90\% of the UEs in the cell obtain a certain average rate. The rationale behind considering these satisfaction conditions is to guarantee each single UE's  minimum rate while at the same time improving the total rate of the cell.

To enable a fair comparison, the set of players and the corresponding set of actions in the proposed satisfaction based MM approach are the same as in the MAB based MM approach. The utility function of player $k$ at time $t_n$ is defined as the load according to equation \eqref{eq:load}.
In the satisfaction based learning approach, the actions are selected according to a probability distribution $\boldsymbol{\pi}_{k}(t_n)=[\pi_{k,1}(t_n),\ldots,\pi_{k,|\mathcal{A}_k|}(t_n)]$. Hereby, $\pi_{k,j}(t_n)$ is the probability with which BS $k$ chooses its action $a_{j(k)}(t_n)$ at time $t_n$. 
The following learning steps are performed in each learning iteration:

\begin{enumerate}
\item In the first learning iteration $t_n=1$  the probability of each action is equal and an action is selected randomly. 
\item In the following learning iterations $t_n>1$, the player changes its action selection strategy only if the received utility does not satisfy the cell, i.e. if the satisfaction condition is not fulfilled. 
\item If the satisfaction condition is not fulfilled, the player $k$ selects its action $a_{j(k)}(t_n)$ according to the probability distribution $\boldsymbol{\pi}_{k}(t_n)$. 
\item Each player $k$ receives a reward $\phi_{k,\text{tot}}(t_n)$ based on the selected actions.
\item The probability $\pi_{k,j}(t_n)$  of action $a_{j(k)}(t_n)$ is updated according to the linear reward-inaction scheme:
\vspace{-0.2cm}
\begin{align}
\begin{split}
\pi_{k,j}(t_n) & = \pi_{k,j}(t_n-T_{\rm s} ) + \lambda \cdot b_k(t_n)\cdot \\
& \biggl( \mathds{1}_{a_{j(k)}(t_n) = a_{i(k)}(t_n)}  -\pi_{k,j}(t_n-T_{\rm s} )\biggr),
\end{split}
\end{align}
\vspace{-0.1cm}
\noindent whereby $\mathds{1}_{a_{j(k)}(t_n) = a_{i(k)}(t_n)}=1$ for the selected action and zero for the non-selected actions and $b_k(t_n)$ is defined as follows:

\begin{equation}b_k(t_n)=\dfrac{u_{k,\text{max}} + \phi_{k,\text{tot}}(t_n)- u_{k,\text{min}}}{2\cdot u_{k,\text{max}}},\end{equation}
with $u_{k,\text{max}}$ being the maximum rate in case of single-UE and $u_{k,\text{min}}= \frac{1}{2}\cdot u_{k,\text{max}}$. Hereby, $\lambda = \frac{1}{100\cdot t_n +T_{\rm s} }$ is the learning rate. 
\end{enumerate}

\section{Simulation Results}
The scenario used in the system-level simulations is based on configuration
\#4b HetNet scenario in~\cite{3GPP36814}. Simulations are performed with the picocell deployment based modified version of the system level simulator presented in~\cite{Simsek}. We consider a macrocell consisting of three sectors, an inter-side distance of 500~m, and $P = \{1,2,3\}$ pico BSs per macro sector,  randomly distributed within the macrocellular environment. In each macro sector, $U = 30$ mobile UEs are randomly dropped within a 60~m radius of each pico BS. The rationale behind dropping all UEs around pico BSs is to obtain a large number of handover within a short time in order to avoid large computation times due to the complexity of our system level simulations. Each UE $i(k)$ has a randomly selected velocity $v_{i(k)}$ of ${\mathcal{V}= \{3;30; 60; 120\}\text{ km/h}}$  and a random direction of movement within $[0;2\pi]$, so that both macro-to-pico and pico-to-macro handover may occur. We consider fast-fading and shadowing effects in our simulations that are based on 3GPP assumptions~\cite{3GPP36814}. 
\begin{figure}
	\centering
		\includegraphics[width=0.45\textwidth]{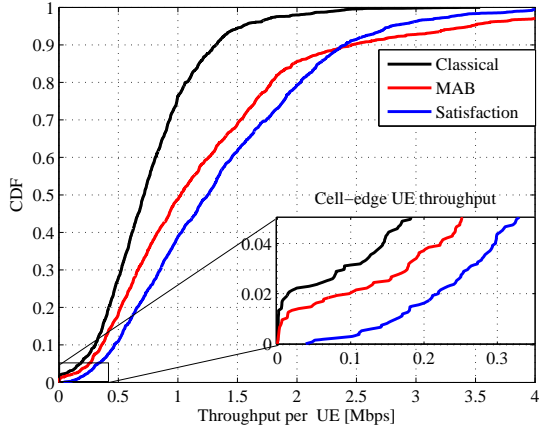}
	\caption{CDF of the UE throughput for 30 UEs and 1 pico BS per macrocell and TTT = 480 ms.}
	\label{fig:CDFofUEThroughput}
\end{figure}
To compare our results with other approaches we consider a baseline MM approach as defined in ~\cite{3GPP36839}. The UE performs RSRP measurements over one subframe every 40 ms and reports this value. The Layer 1 filtering averages the reported RSRP values every 200 ms to filter out fast fading effects. This value is further averaged through afirst-order ifinite impulse response (IIR-)filter which is known as a Layer 3 filter. A handover is then triggered if the Layer 3 filtered handover measurement meets the handover event entry condition in \eqref{eq:HOcondition}.  
A UE is handed over to its target cell after TTT. For the baseline MM approach, we consider proportional fair based scheduling, with no information exchange between macro and pico BSs. This baseline approach is referred to as \emph{classical} HO approach. 

\begin{figure}
	\centering
		\includegraphics[width=0.45\textwidth]{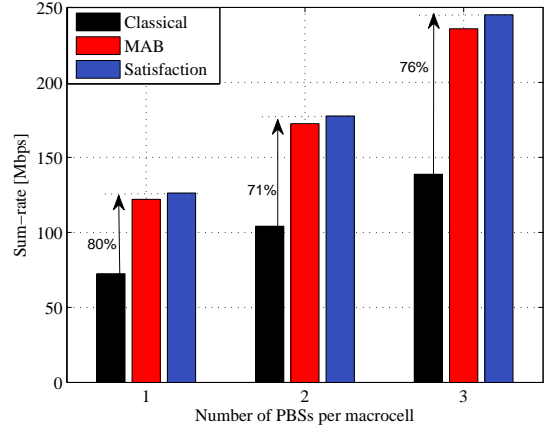}
	\caption{Sum-rate vs. number of pico BSs per macrocell  and TTT = 480 ms.}
	\label{fig:SumRateVsPBSno}
\end{figure}

Fig. \ref{fig:CDFofUEThroughput} depicts the cumulative distribution function (CDF) of the UE throughput for the \emph{classical}, \emph{MAB} and  \emph{satisfaction} based MM approaches. Compared to the classical approach,  MAB and satisfaction based approaches lead to an improvement of 43\% and 75\% in average (50-th~\%), respectively. Hence, the satisfaction based approach outperforms the other MM approaches in terms of average UE throughput. In case of  the cell-center UE throughput, which is defined as the {95-th~\%} throughput, the opposite behavior is obtained. In this case an improvement of 124\% and 80\% is achieved for the MAB and satisfaction based approaches, respectively. The reason is that the satisfaction based MM approach only aims at satisfying the network in terms of rate and does not update its learning strategy once satisfaction is achieved. The MAB based approach on the other hand aims at maximizing the network performance, which is reflected in the improved cell-center UE throughput. The gains of the proposed MM approaches are also reflected in the cell-edge UE throughput, which is zoomed in Fig. \ref{fig:CDFofUEThroughput}. Here, the MAB and satisfaction based approaches yield 39\% and 80\% improvement compared to the classical approach.

\begin{figure}[htb!]
	\centering
		\includegraphics[width=0.45\textwidth]{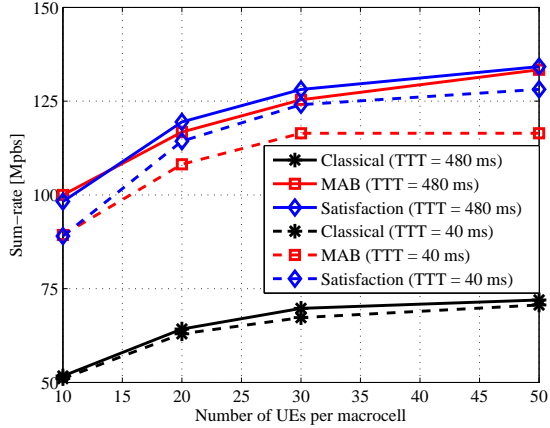}
	\caption{Sum-rate vs. number of UEs per macrocell with 1 pico BS and TTT = 40 ms and TTT = 480 ms.}
	\label{fig:SumRateVsLoad}
\end{figure}

To compare the performance of the proposed approaches for different number of picocells per macrocell, Fig. \ref{fig:SumRateVsPBSno} plots the sum-rate versus number of pico BSs per macrocell. For different number of pico BSs the proposed MM approaches yield gains of around 70\%-80~\% for TTT = 480 ms. In Fig.~\ref{fig:SumRateVsLoad}, the performance of the sum-rate versus UE density per macrocell is depicted for TTT = 40 ms and TTT = 480 ms.  In both cases, the classical approach yields very low rates, while the proposed approaches lead to significant improvement of up to 81~\% for TTT = 40 ms and 85~\% for TTT = 480 ms and converge to a significantly larger sum-rate than the classical approach.

\begin{figure}[htb!]
	\centering
		\includegraphics[width=0.45\textwidth]{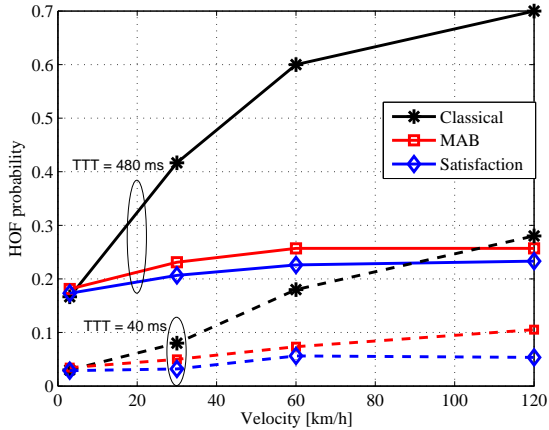}
	\caption{HOF and ping pong probability for 30 UEs and 1 pico BS per macrocell and TTT = 40 ms and TTT = 480 ms.}
	\label{fig:HOFprobability}
\end{figure}
Besides the gains in terms of rate, our proposed learning based approaches yield also improvements in terms of HOF probability as depicted in Fig. \ref{fig:HOFprobability}. For the HOF performance evaluation, we modify our simulation settings by setting the same velocity for each UE. Compared to the classical MM approach, the proposed methods yield the same HOF probability for UEs at 3 km/h speed. For higher velocities in which more HOF is expected, the HOF probability obtained by the proposed approaches  
is significantly lower than in case of classical MM.

\begin{figure}
	\centering
		\includegraphics[width=0.45\textwidth]{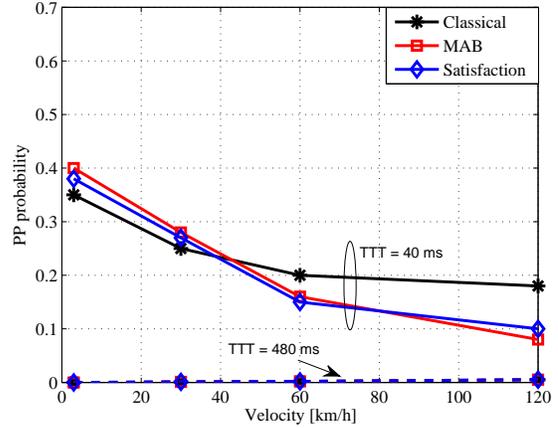}
	\caption{PP probability for 30 UEs and 1 pico BS per macrocell and TTT = 40 ms and TTT = 480 ms.}
	\label{fig:PPprobability}
\end{figure}

The PP probability is depicted in Fig. \ref{fig:PPprobability}. For TTT~=~${40~\text{ ms}}$, all MM methods yield very similar PP probabilities for lower velocities while this probability is decreased for higher velocities. This slope is aligned with the results presented in~\cite{3GPP36839}. However, for high velocity UEs, the PP probability of the proposed MM approaches is half of the PP probability obtained for the classical MM approach which shows a significant improvement. The rationale behind this is that both tiers perform CRE for load balancing, i.e. if one cell tries to extend its coverage/handover a UE the other cell may prevent this handover by extending its coverage, too. In case of TTT~=~${480~\text{ ms}}$ almost no PPs are observed. 
\vspace{-0.3cm}
\section{Conclusion}
We propose two learning based MM approaches and a history based context-aware scheduling method for HetNets. The first learning approach is based on MAB methods and aims at system performance maximization. The second learning method aims at satisfying each cell and each UE of a cell based on satisfaction based learning. System level simulations demonstrate the performance enhancement of the  proposed approaches compared to the classical MM method. While up to 80\% gains are achieved in average for UE throughput, the HOF probability is reduced up to a factor of three by the proposed learning based MM approaches.

\end{document}